\documentclass[reprint,superscriptaddress,showpacs,preprintnumbers,amsmath,amssymb,aps,floats,prl]{revtex4-1}
\usepackage{graphicx}
\usepackage{dcolumn}
\usepackage{bm}
\usepackage{natbib}
\usepackage{xcolor}
\begin{document}

\title{Chiral photodetector based on GaAsN}

\author{R. S. Joshya}
\author{H. Carr\`ere}
\affiliation{Universit\'e de Toulouse, INSA-CNRS-UPS, LPCNO, 135 Avenue Rangueil, 31077 Toulouse, France}
\author{V. G. Ibarra-Sierra}
\author{J. C. Sandoval-Santana}
\affiliation{Area de F\'isica Te\'orica y Materia Condensada, Universidad Aut\'onoma Metropolitana  Azcapotzalco,
Av. San Pablo 180, Col. Reynosa-Tamaulipas, 02200 Cuidad de M\'exico, M\'exico}
\author{V. K. Kalevich}
\author{E. L. Ivchenko}
\affiliation{Ioffe Physical-Technical
Institute, 194021 St. Petersburg, Russia}
\author{X. Marie}
\author{T. Amand}
\affiliation{Universit\'e de Toulouse, INSA-CNRS-UPS, LPCNO, 135 Avenue Rangueil, 31077 Toulouse, France}
\author{A. Kunold}
\affiliation{Area de F\'isica Te\'orica y Materia Condensada, Universidad Aut\'onoma Metropolitana  Azcapotzalco,
Av. San Pablo 180, Col. Reynosa-Tamaulipas, 02200 Cuidad de M\'exico, M\'exico}
\author{A. Balocchi}
\email{andrea.balocchi@insa-toulouse.fr}
\affiliation{Universit\'e de Toulouse, INSA-CNRS-UPS, LPCNO, 135 Avenue Rangueil, 31077 Toulouse, France}

\begin{abstract}
The detection of light helicity is key to several research and industrial applications from drugs production to optical communications. However, the direct measurement of the light helicity is inherently impossible with conventional photodetectors based on III-V or IV-VI semiconductors, being naturally non-chiral. The prior polarization analysis of the light by a series of often moving optical elements is necessary before light is sent to the detector.  A method is here presented to effectively give to the conventional dilute nitride GaAs-based semiconductor epilayer a chiral photoconductivity in paramagnetic-defect-engineered samples. The detection scheme relies on the giant spin-dependent recombination of conduction electrons and the accompanying spin polarization of the engineered defects to control the conduction band population via the electrons' spin polarization. As the conduction electron spin polarization is, in turn, intimately linked to the excitation light polarization, the light polarization state can be determined by a simple conductivity measurement.
This effectively gives the GaAsN epilayer a chiral photoconductivity capable of discriminating the handedness of an incident excitation light in addition to its intensity. This approach, removing the need of any optical elements in front of a non-chiral detector, could offer easier integration and miniaturisation. This new chiral photodetector could potentially operate in a spectral range from the visible to the infra-red using (In)(Al)GaAsN alloys or ion-implanted nitrogen-free III-V compounds.
\end{abstract}
\maketitle
\section{Introduction}
In nature, the animal kingdom has developed the ability to detect or reflect circular polarized light (CPL) as an added modality
 for enhanced vision, better navigation or ``private communications'' \cite{Rossel1986,Cronin2003,Tsyr2008}.
The characterization of light polarization states is also essential in a variety of reasearch and industry fields ranging from
pharmaceutics drugs circular dichroism~\cite{Purdie1996}, polarimetric enhanced vision and imaging~\cite{Khorasaninejad2006}, spectroscopic material characterization, medical and scientific diagnosis~\cite{Greenfield2006,Sparks7816} and optical 
communications~\cite{Farshchi2011}. The development of a simple, efficient and integrable chiral 
photodetector could offer improvements in actual uses but also favour novel device design and applications inspired by natural 
vision systems.\\
The inherent difficulty in discriminating the circular polarization states of light resides in the lack of chirality in 
materials commonly employed in optoelectronics. Common (non-chiral) photodetector are thus used in combination with a set of
moving, and often bulky optics such as quarter waveplates and polarizers~\cite{Berry1977}.
Thus, conventional systems make it a challenge to construct robust, miniature and integrated devices. A direct detection of the 
CPL state is a desirable improvement offering advantages in term of robustness and integrability in addition to a simplified 
technology. Albeit ultra-compact optical elements for manipulating circularly polarized are actively 
investigated~\cite{Yu2012,Gansel2009}, the realization of a simple, 
room temperature ``single-measurement'' circular polarization degree photo-detector remains elusive.
\begin{figure}[!h]
\centering
\includegraphics[width=\linewidth]{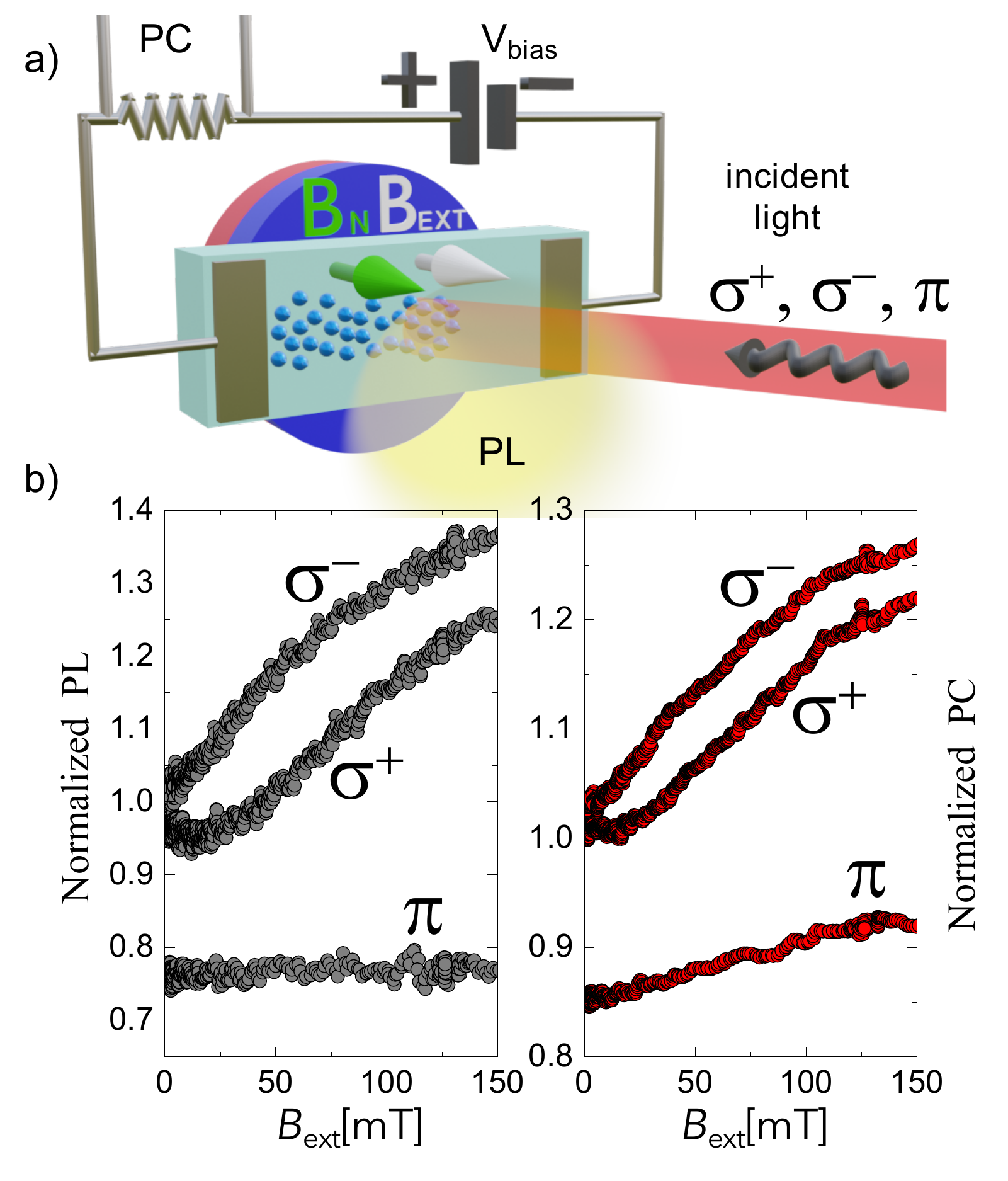}
\caption{ a) Sketch of the experimental configuration used to simultaneously measure the photoluminescence (PL) and the photoconductivity (PC). $B_{\rm N}$ and $B_{\rm ext}$ indicate respectively the induced nuclear magnetic field and the external magnetic field created by a permanent magnet in the configuration when the magnetic fields are parallel (see text).  b)  Variation of PL and PC signals as a function of the external magnetic field for different light polarizations. $\lambda_{\mathrm{exc}}$=852 nm, $P_{\mathrm{exc}}$=20 mW.}
\label{fig:Fig1} 
\end{figure}
Different approaches have put forward plasmonic or CMOS-like nanostructures~\cite{meta_1,meta_2,CMOS} or organic chiral 
molecules~\cite{Schulz2019,Yang2013}. These devices often suffer from limited acceptance angle and with the use of multiple optical elements 
it makes technically challenging to realise miniature and integrated CPL detectors. 
Recent measurements on quantum-dots-based hybrid devices have shown a photocurrent helicity asymmetry of 0.4\%  at 
4 K~\cite{Cadiz2020}, 2\% ($T$=4 K) in a similarly designed quantum wells hybrid ferromagnetic/semiconductor structures or again  5.9\% in a 
Ge-based  device ($T$=4 K)~\cite{Rinaldi2012}.  Better results have been reported  in chiral plasmonic metamaterials. However, these are absorption devices lacking the direct electrical detection~\cite{Hovel2008}. Very recently, 2D WTe$_{2}$ has been used by exploiting the spin photogalvanic effect~\cite{Ji2020}. 
Each of these approaches tackles one or more of the desired characteristics, but suffers some drawbacks such as a narrow spectral range or they do require multiple, complex technological processing steps to be fabricated.\\
\begin{figure}
\centering
\includegraphics[width=\linewidth]{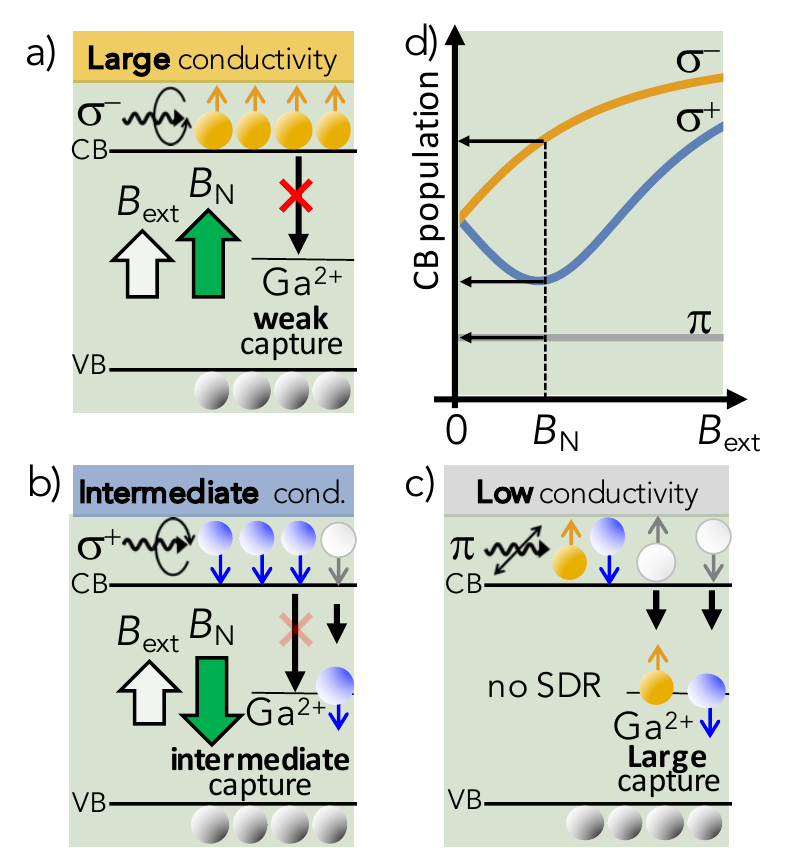}
\caption{a) to c): Schematic view of the SDR capture under different excitation polarizations when a positive (same orientation as the $\sigma^{-}$ photo-generated spins) external magnetic field is applied to the epilayer in Faraday geometry. d) Representation of the corresponding  conduction electron population at dynamical equilibrium.} 
\label{fig:Fig2}
\end{figure}
\begin{figure}
\centering
\includegraphics[width=\linewidth]{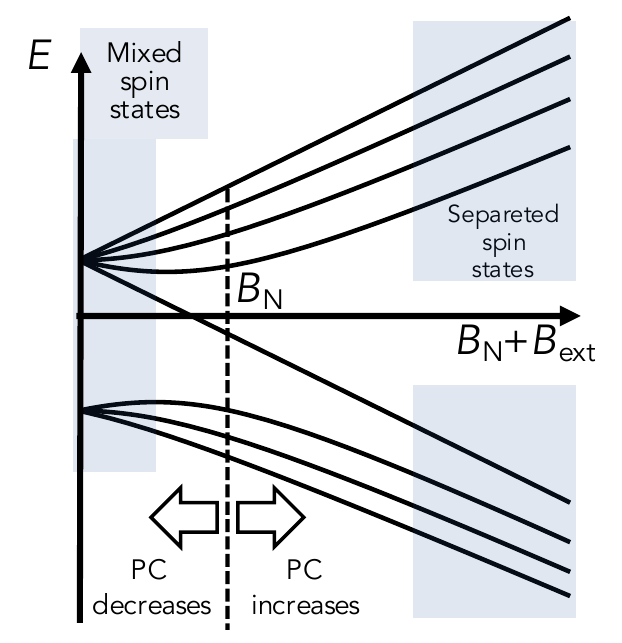}
\caption{Schematic of the hyperfine and Zeeman splitted levels of the electron-defect system as a function of the total magnetic field ($B_{\rm N}+B_{\rm ext}$). The vertical dotted line schematically indicates the value of $B_{\rm N}$ when the external magnetic field is zero. If $B_{\rm ext}$ is decreased or increased from $0$ the defect capture rate will respectively increase or decrease leading respectively to lower or higher PC. The shaded regions near the origin and for high total magnetic field  schematically indicate respectively the regions where the nuclear and electron spin states are fully mixed or substantially independent.} 
\label{fig:Fig3}
\end{figure}

Here we implement the simple solution to this challenge recently proposed in~\cite{kunold-theory} which makes use of a dilute nitride 
GaAsN semiconductor epilayer as the spin photoconductive device capable of determining directly, by a voltage or current 
measurement, the handedness of an incident light. The device operation relies on the spin dependent recombination of conduction electrons and on the hyperfine interaction (HFI) between bound electrons and nuclei on paramagnetic centers
to provide the device with respectively a sensitivity to the degree of circular polarization and its handedness.\\
This semiconductor belongs to the family of dilute nitride materials consisting of a GaAs-based semiconductor where a 
small percentage of N is introduced to induce the formation of Ga$^{2+}$ interstitial paramagnetic defects which display both an 
exceptionally large spin dependent recombination (SDR) for conduction band electrons and a defect nuclear hyperpolarization
 at room temperature~\cite{sdr1,Lombez2005,wang_room-temperature_2009,Kalevich2012}. A very efficient defect recombination is usually an undesirable feature for optoelectronics devices.
 However, as the capture rate is here 
dependent on the spin polarization of the conduction band 
electrons, it provides a way of controlling the conduction band population and thus the conductivity via the spin polarization.
Key to this device workings is the fact that this spin polarization is likewise transferred not only to the defect-bound 
electron but also to the surrounding nuclei via a concomitant action of the hyperfine interaction and dipole-dipole spin 
relaxation~\cite{kunold2020}. This yields a sizeable effective nuclear magnetic field $B_N$  whose 
direction is parallel to the conduction electron spin polarization and thus to the photon helicity.
This phenomenon is however equally efficient for spin up or spin down conduction electron polarization allowing only the 
determination of a circular polarization degree without its handedness. By exploiting in addition the interplay between 
the HFI and a small external magnetic field, the GaAsN epilayer acquires a chiral photoconductivity (PC) capable of discriminating the
 handedness of an incident light in addition to a linearly polarized one as demonstrated below.\\
It is important to note that spin dependent recombination and nuclear polarization can occur on many defects and in various 
materials but typically with a weak efficiency or at cryogenic temperatures. In contrast, in GaAsN-based semiconductors they 
manifest with macroscopic record-high values at room temperature, paving the way to a practical exploitation of the 
phenomenon.
In addition, these paramagnetic defects can be induced to the whole range of GaAs-based alloys from  AlGaAsN to 
InGaAsN~\cite{sdr1}, thus allowing a wide range of wavelength operation from the visible to the infra-red spectral region.
Nitrogen-free gallium-ion implanted semiconductor 
systems have also been shown to display this giant spin-dependent recombination, allowing dissociating the spin dependent 
recombination efficiency and the defect density from the material gap and to offer extended 
device configurations and applicabilities~\cite{Nguyen2013}. Finally, as the the working principle relies on the creation of a spin polarized conduction-band
population, the device's working range can in principle cover a wide spectrum of excitation wavelength as long as a minimum optical orientation in the conduction band can be achieved.
\section{Results and Discussion}
\textbf{Figure 1b} presents the simultaneous measurement of the PL and PC of a GaAsN (2.1\% N) sample for different light polarization states as 
a function of the applied external magnetic field $B_{\rm ext}$ in Faraday geometry for a 20 mW excitation at 852 nm. 
We can see a very similar behaviour for PL and  PC as a function of the magnetic field and light polarization,
 confirming that the same physical phenomena underpin both measurements. 
 From these data two features can be noted: (i) The conduction electron population, as revealed by the PL or PC intensities, depends on 
 the incident light helicity as soon as $B_{\rm ext}\neq$ 0 and for a very small magnetic field strength. This weak field rules out any 
 macroscopic Zeeman effect on the conduction or defect electron states capable of producing a thermal spin population.
(ii) The asymmetry persists for a large range of magnetic fields (the PL and PC intensities under circular polarization
 are expected to eventually become identical for $B_{\rm ext}> 300$ mT~\cite{kunold-theory}). 
These data identify a broad range in external magnetic field where the discrimination of the light polarization can be achieved
 and used in a single-reading electrical detection.
\\
The particular dependence of the PL and PC signals as a function of the external magnetic field for different polarization 
has been successfully interpreted in the framework of an interplay between the hyperfine interaction $\hat{H}_{HFI}=\mathcal{A} \hat{I} \cdot \hat{S}$ (where $\mathcal{A}$
is the hyperfine interaction constant)
  between the Ga$^{2+}$  nuclei (spin $I$=3/2) and the bound
electrons (spin $S$=1/2), and the Zeeman interaction induced by the external magnetic field. The hyperfine interaction reduces the SDR efficiency due to the
mixing of the electron's and nucleus's spins~\cite{Ivchenko2015}, but allows for the flow of angular 
momentum transfer from the conduction band electron to the defect nucleus in combination with the SDR~\cite{ibarra-Sierra2017}.
The way this interplay is exploited can be clarified with the help of \textbf{figure 2} and \textbf{3}: When the spin of the conduction electrons 
shows a preferential orientation (such as under an at least  partially circularly polarized light, figure 2a) the defect resident electron 
acquires the same spin orientation, through the spin dependent recombination, in a few tens of picoseconds~\cite{sdr1}. 
This spin polarization is likewise transferred to the defect nucleus and neighbouring nuclei through a combination of the 
hyperfine interaction and dipole relaxation, yielding an effective nuclear magnetic field $B_{\rm N}$. This spin polarization transfer 
from the conduction band to the defect electron and nuclear spins has a fundamental role: (i) 
by forcing the same  spin  orientation onto the defect resident electron, it prevents any further conduction electron capture thanks
 to the Pauli's  principle. The  capture of the conduction electrons is greatly reduced giving the sample a ``{\bf Large}'' conduction 
 population. Without an external magnetic field this happens with the same efficiency for a right or left circular excitation. 
Instead, under a linearly-polarized excitation (figure 2c) an equivalent density of spin-up and spin-down conduction population is 
photo-generated and no net transfer of spin polarization to the defects can occur. 
The conduction capture is no more spin-dependent and the 
defect can efficiently capture the conduction electrons leading to a ``{\bf Low}" conduction population.
(ii) Under a weak external magnetic field in Faraday geometry, the PL (or PC) intensity under a circularly polarized light 
takes the form of an inverted Lorentzian (figure 2d)~\cite{ibarra-Sierra2017} displaced from the origin in opposite
direction for a right or left circular incident 
polarization due to the presence of the nuclear field $B_{\rm N}$ which adds up to the external field.
Now the system is effectively chiral: When $B_{\rm N}$ is parallel or antiparallel to an applied external magnetic field of comparable 
strength, the HFI in the defect system is proportionally weakened or strengthened with respect to the electron Zeeman effect 
offering a way to proportionally control the SDR efficiency (figure 3). The conduction carrier capture time (figure 2a and b) becomes 
dependent on the total magnetic field $B_{\rm tot}=B_{\rm N}+B_{\rm ext}$. As the sign of $B_{\rm N}$ is directly linked to the 
average conduction electrons spin orientation, the sample conductivity becomes now dependent on the average conduction
electrons spin orientation and therefore on the light polarization state incident on the sample.
\\
In more details, the device works on the following principle: A permanent magnet with a field of 30 to 50 mT is placed on the substrate 
side of the GaAsN epilayer to have the field parallel (or anti-parallel) to the optical axis of the incident light. The magnetic
 field is chosen such that it falls in the optimum region for the discrimination between a circular 
right or left incident polarization~\cite{kunold-theory}, which is of the same order of magnitude as $B_{\rm N}$. 
A constant voltage in the range 0.1 to 4 V is applied to the sample contacts. At this point,  the 
reading of the photoconductivity value reflects the polarization state of the incident light. \textbf{Figure 4a} presents such a 
measurement in the case of a 30 mW incident light at 852 nm under the 50 mT magnetic field. We observe a clear difference 
between the signal measured for a right, left or linear incident polarization, which allows for the determination of
 the light helicity with a simple PC reading. Figure 4b shows how the signal varies when the incident polarization is continuously 
 modified from right 
circular to linear and to left circular by turning the angle of the quarter-wave plate in the incident optical path. This shows in addition that, albeit for a small polarization interval, the device can also determine the 
polarization degree of an incident light in a single measurement. Considering that the root mean square error $RMS=0.0068$ mV as measured from a linear fit of the first plateau signal for a $\sigma^-$ excitation in figure 4a, and the signal difference between a linear and $\sigma^-$ excitation, $\delta=$0.36 mV, we can estimate a measurement of a circular polarization degree with a precision of about 2\% in this proof of concept device.
\begin{figure}
\centering
\includegraphics[width=\columnwidth]{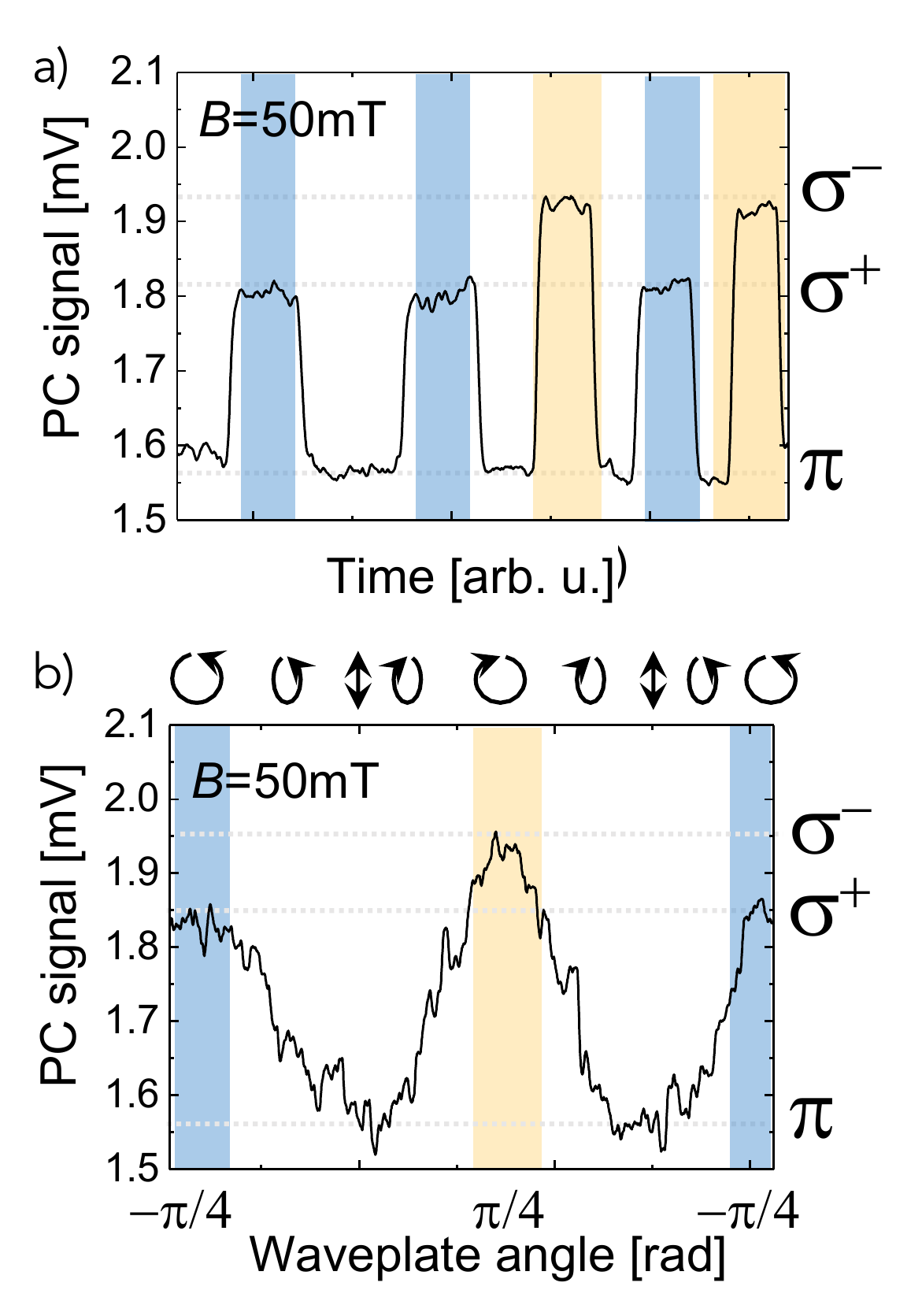}
\caption{(a) Photoconductivity signal  measured at 50 mT for different polarizations of the incident light. b) Photoconductivity 
signal measured as a function of the quarter-wave plate angle with respect to the incoming linearly polarized light excitation.}
\label{fig:device-working} 
\end{figure}
\begin{figure}
\centering
\includegraphics[width=\columnwidth]{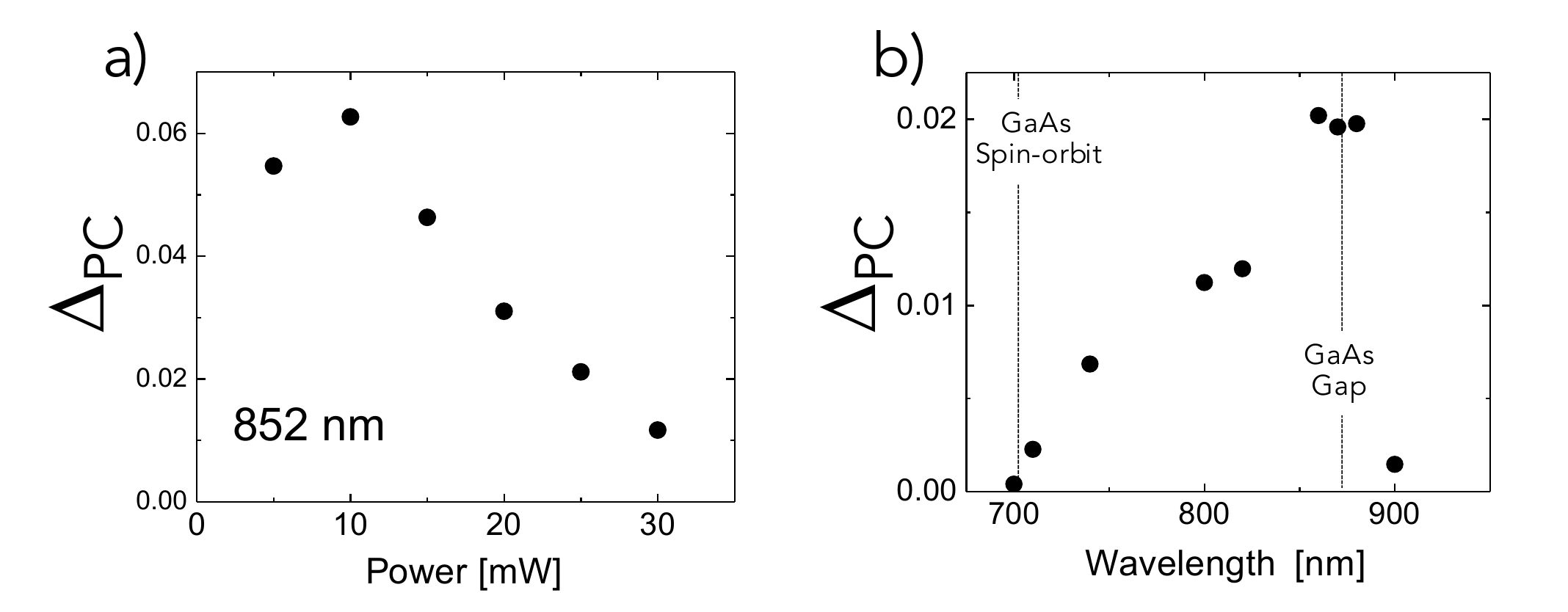}
\caption{a) The PC contrast $\Delta_{PC}$ as a function of the excitation power at 852 nm excitation. b) The PC contrast as a function of the excitation wavelength.}
\label{fig:Fig3}
\end{figure}
We have characterized the PC contrast $\Delta_{PC}=\frac{PC^{-}-PC^+}{PC^{-}+PC^+}$ ($PC^{-/+}$ indicates respectively the 
PC under a $\sigma^-$ or $\sigma^+$ excitation) as a function of the incident wavelength and power. \textbf{Figure 5a} reports the data 
obtained as a function of the SDR ratio at $B_{\rm N}$=0, SDR$_{r}(0)=100\frac{PC^{\sigma}}{PC^{\pi}}$, by varying the incident 
power. The contrast reaches a maximum for a relatively weak powers and 
then monotonically decreases as the power is increased. This behaviour is well accounted for~\cite{kunold-theory} at low power 
by the power threshold for the build-up of the nuclear magnetic field. Increasing the excitation power beyond the optimum leads to a decrease of the SDR, leading
 in return to a decrease of the photocurrent contrast. 
Figure 5b presents the PC contrast as a function of the excitation wavelength. A constant incident power $P_{\rm exc}$= 20 mW 
has been used here for the whole spectral range explored. We can deduce that for this Nitrogen content and for this simple 
photoconductivity configuration, the device can be used over 100 nm range. The hight wavelength limit ($\lambda$= 870 nm, $E$= 1.424 eV) is due to
the change in the absorption coefficient of the device once the excitation energy is smaller than the GaAs gap energy. Exciting with photon energies above
 the GaAs gap allows instead for a larger absorption of the incoming light as this can occur over the thicker GaAs buffer layer and not only on the thin 100 nm GaAsN one, while preserving a substantial conduction electron spin polarization during the thermalisation and
 diffusion into the GaAsN layer~\cite{sipe}. No higher excitation wavelength has been tested. However,
   thanks to the optical selection rules, a PC contrast should in principle be observable for wavelengths up to the GaAsN gap 
   as the conduction spin polarization will increase for excitations closer to the gap, albeit higher powers will be necessary to compensate for the lower absorption in the thin layer. The low wavelength limit ($\lambda$= 704 nm, $E$= 1.76 eV) is instead linked to the onset of the spin-orbit
   band where no or very weak conduction spin polarization can be achieved.  It is important also to consider here that both the 
absorption coefficient and the photogenerated conduction spin polarization are wavelength dependent. As the SDR depends
on the absorbed power and the spin polarization achieved at the bottom of the conduction band, the relation between the PC 
contrast and wavelength is non-linear.

As suggested in~\cite{kunold-theory} an improved version of the device 
could be envisaged in the form of either a balanced photodetector composed of two slabs of GaAsN each subject to opposite
 magnetic fields $\bf{B_1}$ and $\mathbf{B_2}=-\mathbf{B_1}$. A  common voltage source is used to 
drive in both GaAsN  slabs and the measured PCs from the two photodetectors are subtracted. In this configuration, 
the sign and not the magnitude of the resulting current will reveal the polarization state of the incident light.  An alternative solution 
is also proposed where the differential measurement is obtained in a single GaAsN layer subject to an alternating magnetic field
produced by a micro-coil powered by an alternating current (\textbf{figure 6a}). 
In order to make a proof of concept, we have tested a modified version of the second device design where the single GaAsN
is here subject to a constant magnetic field while the incident light has a periodically varying polarization state (\textbf{figure 6b}).
The polarization modulation is obtained by rotating a quarter-wave
 plate at a frequency $f_{\mathrm{\rm rot}}=\omega_{\rm rot}/2\pi$. In this configuration, the polarization state after the waveplate will periodically change 
 from linear to circular right, linear, circular left and back to linear, twice per rotation.
By demodulating the measured photocurrent at a frequency $2 f_{\rm rot}$ using a lock-in amplifier, the resulting signal is proportional to $(i^{\sigma-}  -i^{\sigma+})$, i.e. the difference in the photocurrent signal between a circular right and circular left polarized excitation. 
\begin{figure}
\centering
\includegraphics[width=\columnwidth,angle=0]{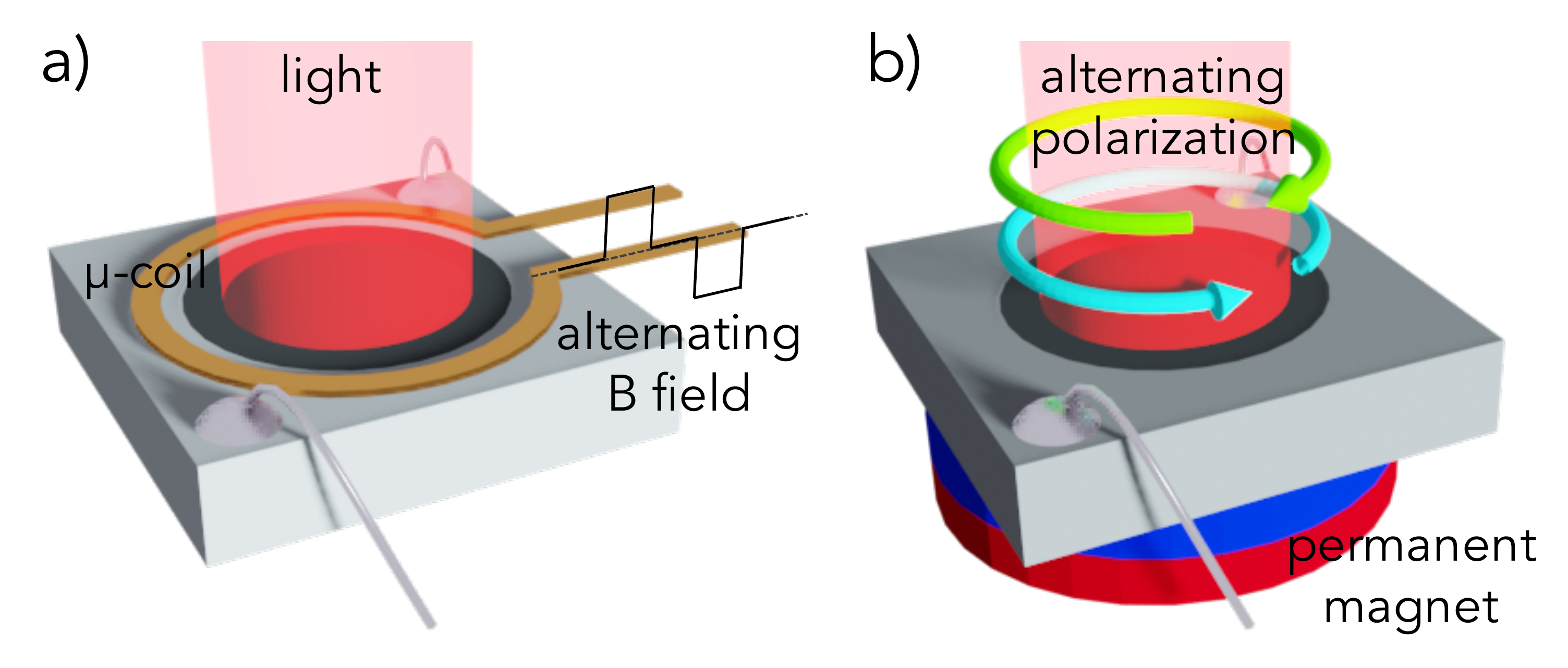}
\caption{a) A representation of one of the differential measurement photodetector designs proposed in~\cite{kunold-theory}. 
A micro-coil provides an alternating positive and negative magnetic field and the light polarization is determined by
 the resulting sign of the measured current by subtracting the measurements obtained under opposite magnetic fields. 
 b) The scheme used here for the proof of concept where a constant magnetic field is used and the device is illuminated with an alternating circular polarization.}
\label{fig:rotating_scheme}
\end{figure}
\begin{figure}
\centering
\includegraphics[width=\columnwidth]{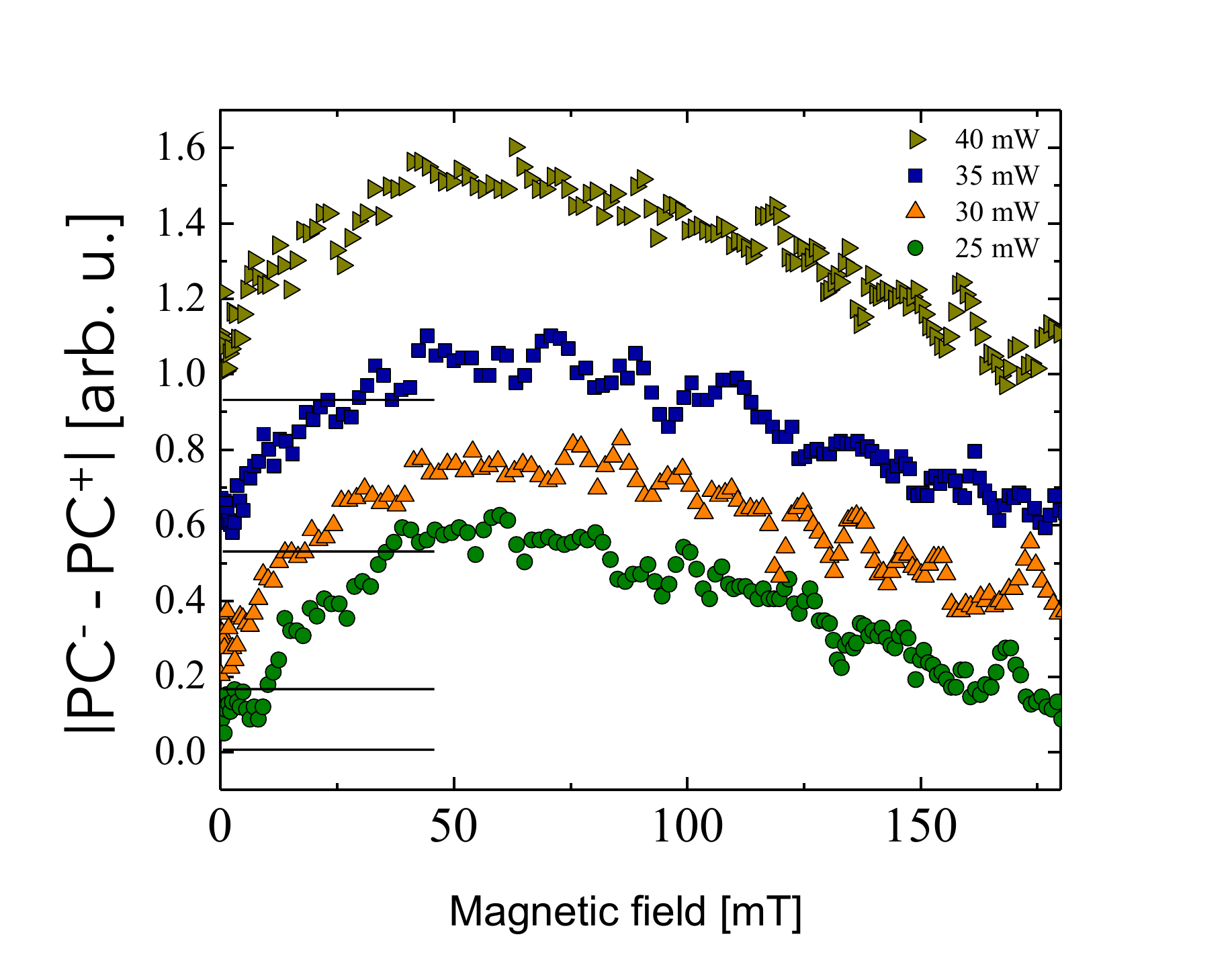}
\caption{The PC  measured using the polarization-modulation technique as a function of the incident power. The curves have been vertically displaced for clarity.}
\label{fig:rotating_scheme}
\end{figure}
\textbf{Figure~\ref{fig:rotating_scheme}} reports the direct measurements of the difference between a circular right and circular left 
incident polarization using the set-up presented above, as a function of the excitation power at 852 nm for 25 - 40 mW and 
 $f_{\rm rot}$= 21 Hz.
We observed that a sizeable differential signal can be obtained for a large range of external magnetic fields. 
In the differential configuration the signal will be different than zero only for circularly  polarized light and the limit to the minimum measurable circular polarization degree will be set mainly by the noise equivalent signal of the reading. 
This is expected to be greatly  improved with a photodiode design instead of a photoconductor one. 
A linear or unpolarized light will give instead a zero reading.\\
\section{Conclusions}
In conclusion, we have shown that leveraging the giant spin dependent recombination and the hyperpolarization of Ga$^{2+}$ interstitial defect in dilute nitrides epilayers allows for the realization of a chiral III-V dilute nitride semiconductor useable for the
 direct electrical measurement of the incident light helicity at room temperature. As the working principle relies on the optical orientation of conduction electrons, the working spectral range could be easily extended for wavelengths up to the material band gap and as long a non zero spin polarization of photoexcited electrons can be achieved.
This paradigm, enabled by the presence of Ga paramagnetic defects in dilute nitrides, can in principle be extended to the whole family of (In)(Al)GaAsN alloys with gaps ranging from the visible to the infrared. The possibility of producing interstitial defects by implantation in N-free alloys~\cite{Nguyen2013} could also provide further application possibilities.
\section{Methods}
The sample studied for the proof of principle demonstration consists of a $100$ nm thick
GaAs$_{1-x}$N$_x$ epilayer ($x=0.021$)  grown by molecular beam epitaxy on a (001) semi-insulating GaAs substrate and
capped with $10$ nm GaAs. The room temperature gap is at 1080 nm. The excitation light was provided by either a 852 nm laser diode  or a continuous wave Ti:Sa laser as the 
tunable wavelength source.
The lasers were focused to a $\approx$ 100 $\mu$m diameter spot (FWHM), in between two Ag electrodes deposited onto the 
sample surface about 1 mm apart ({figure 1a}).  The  laser light polarization was controlled by a Glan-Taylor polarizer followed by a quarter 
wave plate. The laser intensity was modulated by a mechanical chopper at 170 Hz. The sample photoconductivity 
(PC) was measured synchronously using a lock-in amplifier from the voltage drop at the terminals of a 1 M$\Omega$ load 
resistor placed in series with the sample.  We have carefully focused the laser in order to avoid any spurious effects related to
the contact illumination. A constant voltage has been applied between the sample electrodes to perform the PC reading. 
The photoluminescence (PL) intensity was simultaneously measured with the same lock-in technique by recording its total 
intensity, filtered of the laser scattered light and substrate contribution using a series of long-pass optical filters, and integrated by 
an InGaAs photodiode. A permanent Neodymium magnet has been used to apply a magnetic field in the Faraday geometry and
the field strength on the sample surface has been varied by changing the magnet's distance from the sample. All the experiments 
 were performed at room temperature.\\

\medskip

\bibliographystyle{MSP}
\bibliography{biblio}
%

\end{document}